\documentclass[
    ,final            
  ]
  {aipproc}

\layoutstyle{6x9}

\begin{document}

\def\simg{\mathrel{%
      \rlap{\raise 0.511ex \hbox{$>$}}{\lower 0.511ex \hbox{$\sim$}}}}
\def\siml{\mathrel{%
      \rlap{\raise 0.511ex \hbox{$<$}}{\lower 0.511ex \hbox{$\sim$}}}}

\def\epsi{\epsilon_i} \def\epsB{\epsilon_B} \def\cm3{\,\rm cm^{-3}}
\def\E{{\cal E}} \def\Mi{{\cal M}_i}

\title{On the Shallow Decay of Some GRB Afterglows}

\author{A. Panaitescu \& P. Kumar}{ address={Department of Astronomy, University of Texas at Austin} }

\begin{abstract}
  Half of the radio afterglows for which there is a good temporal coverage
 exhibit after 10 days from the burst a decay which is shallower than at optical 
 frequencies, contrary to what is expected within the simplest form of the standard 
 model of relativistic fireballs or jets. We investigate possible ways to decouple 
 the radio and optical decays. First, the radio and optical emissions are assumed
 to arise from the same electron population and we allow for either a time-varying
 slope of the power-law distribution of electron energy or for time-varying
 microphysical parameters. Then we consider two scenarios where the radio and
 optical emissions arise in distinct parts of the GRB outflow, either because 
 the outflow has an angular structure or because there is a long-lived reverse
 shock. We find that the only the last scenario is compatible with the observations.

\end{abstract}

\maketitle

{\bf The anomalous radio afterglows.}
 The radio emission of all well-observed GRB afterglows decays after about day 10. 
This is consistent with what is expected in the standard fireball model for GRB afterglows. 
In this model, the characteristic synchrotron frequency $\nu_i$ at which electrons with 
the typical post-shock energy $e_i = \epsi m_p c^2 \Gamma$ ($\Gamma$ being the fireball 
Lorentz factor) radiate is
\begin{equation}
 \nu_i \sim 30 \; \left( \frac{\E}{10^{53}\, {\rm ergs}} \right)^{1/2} 
                      \left( \frac{\epsi}{0.03} \right)^2 
                      \left( \frac{\epsB}{10^{-3}} \right)^{1/2} 
                      \; \left( \frac{t}{10\,{\rm d}} \right)^{-3/2}\; {\rm GHz} \;
\label{nui}
\end{equation}
where $\E$ is the fireball's kinetic energy per solid angle and the parameters $\epsi$
and $\epsB$ quantify the fraction of the post-shock energy imparted to electrons\footnote{
 For a power-law distribution of electron energies $dN/de \propto e^{-p}$ at $e > e_i$,
 the total electron energy is $[(p-1)/(p-2)]\times \epsi$ for $p > 2$.}
and to the magnetic field. If the outflow is collimated, the evolution of $\nu_i$ becomes
faster, $\nu_i \propto t^{-2}$, after the "jet-break" time when the jet start expanding 
laterally.
 
 Equation (\ref{nui}) indicates that, for reasonable afterglow parameters, the injection
frequency $\nu_i$ is expected to cross the radio domain around 10 days, after which 
the radio afterglow should decay as $F_r \propto t^{-\alpha_r}$ with $\alpha_r = (3p-1)/4$ 
for a fireball interacting with a wind medium, $\alpha_r = (3p-3)/4$ for a fireball decelerated 
by a uniform medium, and $\alpha_r = p$ for jet spreading laterally. In the standard afterglow 
model, the optical light-curve decay is expected to be the same, apart from a difference 
$|\alpha_o -\alpha_r| = 1/4$ arising when the cooling frequency ($\nu_c$) is below the optical 
domain.

 For five radio afterglows (970508, 980329, 980703, 000418, 021004), the radio and
optical decay indices are consistent with this basic "prediction" of the fireball model,
but it is not so for the other five well-monitored radio afterglows (991208, 991216,
000301, 000926, 010222), which are shown in Figure 1 . For these cases, it is natural
to investigate first if the difference between the radio and optical decay indices could
be caused by that the injection frequency $\nu_i$ remains above the radio domain (of
typical frequency $\nu_r \sim 10$ GHz) until the last radio measurements\footnote{ 
  That radio afterglows decay after 10 days indicate that the self-absorption frequency 
  is below the radio domain, thus it cannot account for this difference}
, usually around 100 days after the burst. According to equation (\ref{nui}), this requires 
electron and/or magnetic field parameters close to equipartition ($\epsi, \epsB \simg 0.1$), 
particularly if the jet-break time occurs at around 1 day, as is the case for the afterglows 
991216, 000301, 000926 and 010222 (see the break exhibited by their optical light-curves). 

\begin{figure}
  \includegraphics[width=15cm]{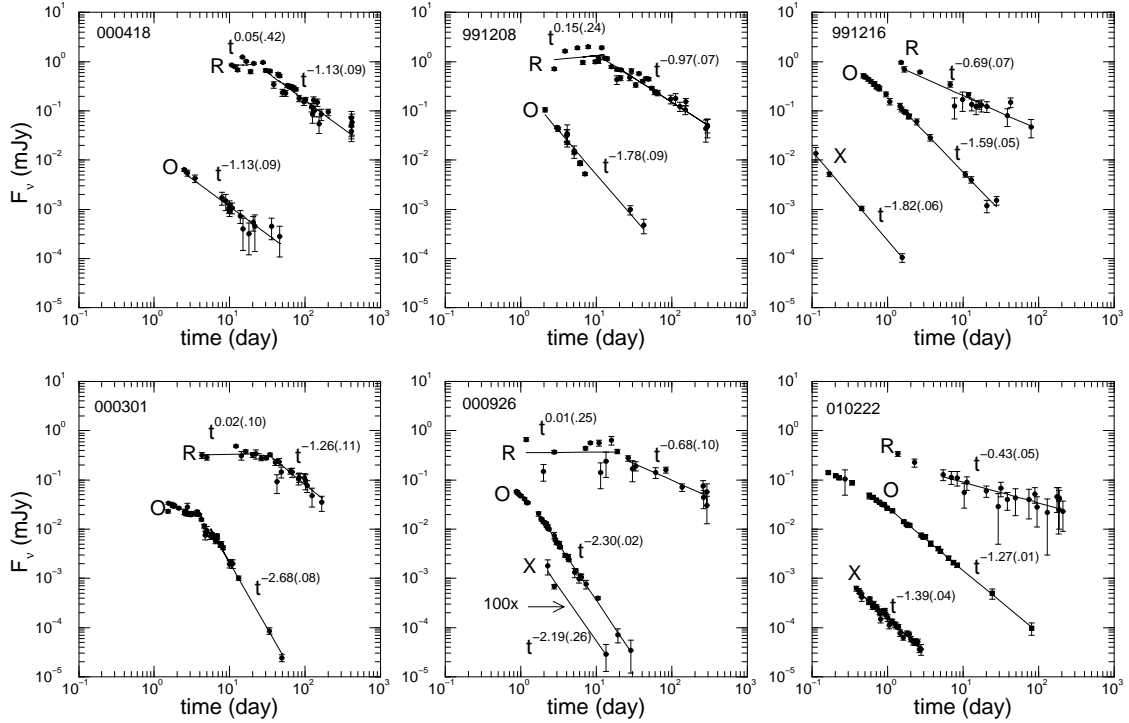}
  \caption{Radio (8 GHz), Optical ($5 \times 10^{14}$ Hz), and X-ray (5 keV) light-curves
   for five anomalous GRB afterglows whose radio light-curves decay slower than in the optical
   and for the afterglow 000418, for which the decays seen in these domains have the same 
   indices, as expected in the simplest version of the standard fireball/jet afterglow model.
   Power-law fits are indicated for each frequency range, errors are given in parentheses.}
\end{figure}

 For $\nu_r < \nu_i$ there are three relevant cases which yield a decaying radio emission.
If the GRB ejecta is spherical (or a sufficiently wide jet), the radio light-curve decays
only if the circumburst medium is a wind ($r^{-2}$ density profile) and if $\nu_r  <\nu_c < 
\nu_i$, in which case $\alpha_r = 2/3$, as observed for the afterglows 991216 and 000926. 
If the GRB ejecta is collimated, a decaying radio light-curve can be obtained either for
$\nu_i < \nu_c$ or for $\nu_c < \nu_i$. In the former case $\alpha_r = 1/3$, close to the
value observed for the afterglow 010222, while the latter case yields $\alpha_r = 1$, as
seen in the afterglow 991208. Despite this nice agreement between the expected
and observed radio decay indices for $\nu_r < \nu_i$, none of the above cases actually
work well. The first and third case require dense external media (homogeneous or wind-like)
to maintain $\nu_c < \nu_i$ until about 100 days, leading to self-absorbed sources at radio 
frequencies even at day 10, which is inconsistent with the decaying radio light-curves,
and to a synchrotron flux at $\nu_i$ above 100 mJy at 100 days, which is 1,000 times 
larger than the radio flux observed at that time. The second case leads to very tenuous
external media, which are rather inconsistent with those expected for a massive star
GRB progenitor.

 Thus we have to investigate departures from the standard afterglow model in its
simplest form that could account for the shallow radio decays observed for the above five
anomalous afterglows. Since the $\nu_r < \nu_i$ case
cannot explain all the properties of the radio afterglow emission, we will consider
that the injection frequency $\nu_i$ is below the radio domain when the radio decay
is observed. In fact, the passage of $\nu_i$ through this domain is the most natural
cause for the onset of the radio decay seen at about 10 days in the afterglows 991208,
000301, 000926 and 010222, as well as in other afterglows whose radio and optical
decay indices are consistent with the fireball model expectations.

 These departures fall in two categories. In the first, we assume that the radio and 
optical emissions arise from the same region of the GRB fireball. If the same electron
population gives both the radio and optical afterglows, then different decay indices
can be obtained either if the slope $p$ of the electron energy distribution is time-varying
or if there is a spectral break between these two domains. In the second category, 
we assume that the radio and optical emissions arise from different fireball regions, 
as could be the case with a structured fireball or a long-lived reverse shock. 
We note that, in the scenarios that we describe below, involving time-varying afterglow 
parameters, the break seen in the optical light-curve of most afterglows shown in Figure 1 
may arise from a rapid variation of the parameters under investigation and not necessarily 
from the tight collimation of the GRB outflow.

{\bf Electron distribution with time-varying slope.} 
 We consider that, at all times of interest (1-100 days after the burst), the electron
distribution injected by the forward shock is a power-law of exponent $-p$ that varies 
in time\footnote{
 An alternative scenario for the $\alpha_o -\alpha_r$ index difference of the anomalous 
 afterglows is an electron distribution that is not a power-law and whose shape does not 
 change in time. This scenario is largely unconstrainable with current observations.}
. It is evident that, in this case, the radio and optical light-curves cannot both be
power-laws in time. Given that optical measurements have smaller uncertainties, we can 
determine the evolution of the electron index $p$ that yields a power-law optical light-curve
and test the consequences of that evolution on the afterglow emission at other frequencies. 

 It can be shown that, in order to obtain $\alpha_r < \alpha_o$, the index $p$ must increase 
in time, asymptotically approaching the value that the observed $\alpha_r$ required if the 
index $p$ were constant in time. Thus the optical afterglow spectrum should soften in time.
For such a behavior of $p$, the radio light-curve should steepen in time, while the $X$-ray
light-curve should decay faster that in the optical and should flatten in time. In general,
these features are not quantitatively consistent with the multiwavelength observations of 
the anomalous afterglows: the optical spectrum of 991208 appears to harden in time while
the decays of the $X$-ray light-curve of 991216 (during day 1) and of the radio light-curves 
of 000926 and 010222 (at 100 days) are much shallower than expected. Only for 000301 the radio 
light-curve and optical spectrum are consistent with the consequences of an evolving electron 
distribution index.\footnote{
 There are no $X$-ray measurements for the afterglow 000301 to further test this scenario}

{\bf Variable electron and magnetic field parameters.}
 As the self-absorption and injection break frequencies must be below the radio domain
when the radio light-curves decay, the only remaining break that could decouple the radio
and optical decay indices must be the cooling frequency $\nu_c$. However its evolution 
must be much faster than that for a constant magnetic parameter $\epsB$, to account for 
the magnitude of the observed $\alpha_o -\alpha_r$. Since we want, in fact, to explain
not just the difference $\alpha_o -\alpha_r$, but the observed values of $\alpha_r$
and $\alpha_o$, we must also allow for the electron energy parameter $\epsi$ to be
time-varying. We shall also consider that the fireball's kinetic energy per solid angle
$\E$ (or the kinetic energy for a jet) is time-varying, either because of radiative losses
or an energy injection in the fireball through some less relativistic ejecta which catch-up 
with the leading edge of the ejecta (which is decelerated by the interaction with the 
circumburst medium).

 Because the quantities pertaining to the fireball dynamics are power-laws in the observer
time $t$, it is then natural to restrict our attention to time-varying $\epsi$, $\epsB$, 
and $\E$ that evolve as power-laws with $t$: $\E \propto t^e$, $\epsi \propto t^i$, $\epsB 
\propto t^b$. Radiative losses yield $e \geq -3/7$ for a fireball interacting with a 
homogeneous medium, $e \geq -1/3$ if the medium is wind-like, and $e \geq -3/5$ for a
spreading jet and any type of external medium. Using standard equations for the afterglows 
spectral characteristics, we can calculate the decay indices for the radio and optical
light-curves as functions of the parameters $e$, $i$ and $b$ and use the observed $\alpha_r$
and $\alpha_o$ to constraint them.

 In this way it can be shown that, if there is no energy injection ($e \leq 0$),
then $\epsi$ must decrease in time while $\epsB$ must increase so fast that the magnetic
field strength $B \propto \epsB^{1/2}$ is constant or increases in time. This
rather extreme requirement is somewhat alleviated if the external medium density increases
with radius, as such a density profile leads to a faster evolution of the cooling frequency.
If there is an energy injection, leading to $e \geq 0$, then there are solutions with 
constant $\epsB$, however the peak flux and injection frequency evolutions implied by this
scenario are in strong conflict with those inferred from the radio emission of the
afterglow 991208. These conclusions apply to both a spherical fireball and
a collimated outflow.

{\bf Structured outflow.} 
If the outflow has a non-uniform angular distribution of the kinetic energy per solid 
angle $\E$, it is possible that the optical emission arises predominantly from a core of 
higher $\E$ while the radio afterglow is emitted by a surrounding envelope of lower
$\E$. This possibility is suggested by the dependence of the injection frequency $\nu_i$
on $\E$ (eq. [\ref{nui}]). For simplicity and maximal effect of the outflow structure, 
we consider that both the core and envelope have an uniform distribution of $\E$. 
In this scenario, the break exhibited by the afterglow optical light-curve is caused by 
the collimation of the core, while the onset of the radio decay is due to the passage 
through the radio domain of the $\nu_i$ for the envelope emission. To explain the observed
$\alpha_r$ and $\alpha_o$, we must allow for different electron indices $p$ in the
outflow core and envelope.

 The test that this scenario must pass is that the radio and optical emissions are decoupled.
More specifically, the softening emission from the 
optical core must not overshine in the radio the emission from the envelope and the 
emission from the envelope must not be brighter in the optical than the faster 
decaying emission from the core. The first condition is equivalent to a lower limit on the 
$\nu_i$ frequency for the optical core at the time when the optical light-curve break is 
observed (which is the time when the core edge becomes visible to the observer). With the 
aid of equation (\ref{nui}), this leads to a lower limit on the injection frequency for 
the radio envelope at the time when the radio light-curve begins to decay. For the parameters 
of the radio and optical emission of the five anomalous afterglows, the latter lower limit 
falls invariably above 10 GHz, contrary to what is implied by the observed onset of the radio 
decay. The second condition leads to an upper limit on the cooling frequency for the
envelope emission, which implies very dense external media and a self-absorbed radio emission,
inconsistent with the observed radio decay.

{\bf Reverse shock.}
If there is a continuous inflow of slower ejecta from the GRB progenitor into the leading
edge of the GRB fireball, then the emission from the long-lived reverse shock crossing the 
incoming ejecta could dominate the radio afterglow emission, while the optical afterglow
arises as usually from the forward shock. Given that a collimated outflow undergoing lateral
spreading and delayed energy injection loses angular uniformity, we restrict our attention to a 
spherical outflow, which maintains its isotropy during the injection.

 For simplicity, we parameterize the distribution of ejecta mass with Lorentz factor as a 
power-law, $d\Mi/d\Gamma \propto \Gamma^{-(q+1)}$. The fireball energy can also vary in 
time (e.g. $\E \propto t^e$), as described for a structured outflow\footnote{
 If the energy of the incoming ejecta is dominant, then the exponents $q$ and $e$ are
not independent, but they are so for a negligible energy injection, i.e. $e \siml 0$}
. Just as for the scenario involving variable microphysical parameters, one can calculate 
the radio and optical decay indices as function of the parameters $q$ and $e$, and then 
determine these parameters with the aid of observations. We find that the anomalous radio 
afterglows require $q > 3$ for a homogeneous medium, $q > 4$ for a wind medium, and $e < 1/3$, 
indicating that the incoming ejecta can carry at most the same energy as the initial outflow
energy.

 The reverse-forward shock scenario must also pass the test discussed above for a structured 
outflow, leading to a lower limit on the forward shock $\nu_i$ frequency and an upper limit 
on the reverse shock cooling frequency. Two other constraints can be obtained by requiring 
that the forward shock yields the flux normalization seen in the optical and that the $\nu_i$ 
frequency for the reverse shock crosses the radio domain when the onset of the radio fall-off 
is seen. These four requirements can be converted into constraints on the fundamental afterglow 
parameters $\E$, $\epsi$, $\epsB$, and $n$ (external medium density). For the anomalous 
afterglows we find that reasonable values of these parameters satisfy the observational 
requirements. 

{\bf Note:} More details on the features of the scenarios described here and the
 calculations behind the results presented can be found at {\sl astro-ph/0308273}.

\end{document}